\documentclass[paper,amsfonts,notoc]{JHEP}

\usepackage{epsfig}

\newcommand\sss{\scriptscriptstyle}
\def\hs{\hat s}
\def\mn{{\sss {\rm MN}}}
\def\as{\alpha_{\sss S}}
\def    \Ecut             {\mbox{$E_\perp$}}
\def    \Ecutt            {\mbox{$E_\perp^2$}}
\def    \muf              {\mbox{$\mu_{\sss F}$}}
\def    \muft             {\mbox{$\mu_{\sss F}^2$}}
\def\dy{\Delta y}

\def    \kta              {\mbox{$k_{a\perp}$}}
\def    \ktb              {\mbox{$k_{b\perp}$}}
\def    \ktat             {\mbox{$k_{a\perp}^2$}}
\def    \ktbt             {\mbox{$k_{b\perp}^2$}}
\def\yw{y_{\sss W}}
\def\mw{m_{\sss W}}
 
\def\gsim{\mathrel{\raisebox{-.6ex}{$\stackrel{\textstyle>}{\sim}$}}} 

\def\beq{\begin{equation}}
\def\eeq{\end{equation}}
\def\beqa{\begin{eqnarray}}
\def\eeqa{\end{eqnarray}}

\def\eqn#1{Eq.~(\ref{#1})}

\def\fig#1{Fig.~{\ref{#1}}}

\title{Forward jets and forward $W$-boson production
 at hadron colliders}

\author{J.~R.~Andersen and W.J. Stirling\\
Institute for Particle Physics Phenomenology, 
University of Durham\\ Durham, DH1 3LE, U.K.\\
E-mail: \email{Jeppe.Andersen@durham.ac.uk, W.J.Stirling@durham.ac.uk}}

\author{V. Del Duca\thanks{Rapporteur at
EPS2001, Budapest, Hungary}\\ 
Istituto Nazionale di Fisica Nucleare, Sez. di Torino\\
via P. Giuria, 1 - 10125 Torino, Italy\\
	E-mail: \email{delduca@to.infn.it}}

\author{S. Frixione\thanks{On leave from INFN, Sez. di Genova, Italy}\\
LAPP\\ Chemin de Bellevue, BP 110,
74941 Annecy-le-Vieux CEDEX - France\\
	E-mail: \email{Stefano.Frixione@cern.ch}}

\author{F. Maltoni\\
Department of Physics,
University of Illinois at Urbana-Champaign  \\
Urbana, IL\ \ 61801, USA\\
	E-mail: \email{maltoni@uiuc.edu}}

\author{C.R. Schmidt\\
Department of Physics and Astronomy,
Michigan State University\\
East Lansing, MI 48824, USA\\
	E-mail: \email{schmidt@pa.msu.edu}}

\abstract{
In this talk we give a short review of forward jets and 
forward $W$-boson production at hadron colliders, in view of the extraction
of footprints of BFKL physics. We argue that at Tevatron energies, 
dijet production at large rapidity intervals is still subasymptotic 
with respect to the BFKL regime, thus the cross section is strongly dependent
on the various cuts applied in the experimental setup.
In addition, the choice of equal transverse momentum cuts on the tagging
jets makes the cross section dependent on large logarithms of non-BFKL origin,
and thus may spoil the BFKL analysis.
For  vector boson production in association with two jets, we argue
that the configurations that are kinematically
favoured tend to have the vector boson forward in rapidity.
Thus $W + 2$ jet production lends itself
naturally to extensions to the high-energy limit.}

\begin{document}

In strong-interaction processes characterised by two
large and disparate energy scales, which are typically the squared
parton center-of-mass energy $\hat s$ and momentum transfer 
$\hat t$, with $\hat s\gg \hat t$, the BFKL theory~\cite{Kuraev:1976ge} 
resums the large logarithms $\ln(\hat s/|\hat t|)$.
Over the past years several attempts have been made to predict and detect
footprints of emission of BFKL gluon radiation in strong-interaction 
processes, like in dijet production at hadron colliders at large rapidity
intervals, in forward jet production in DIS and in $\gamma^*\gamma^*$
collisions in double-tag events, $e^+\, e^- \to e^+\, e^- +$ hadrons.
Here we shall review first dijet production at hadron colliders at 
large rapidity intervals, and then consider the production of a
forward $W$-boson in association with two jets. 

\section{Dijet production at large rapidity intervals}

\begin{figure}[htb]
  \centerline{ \epsfig{file=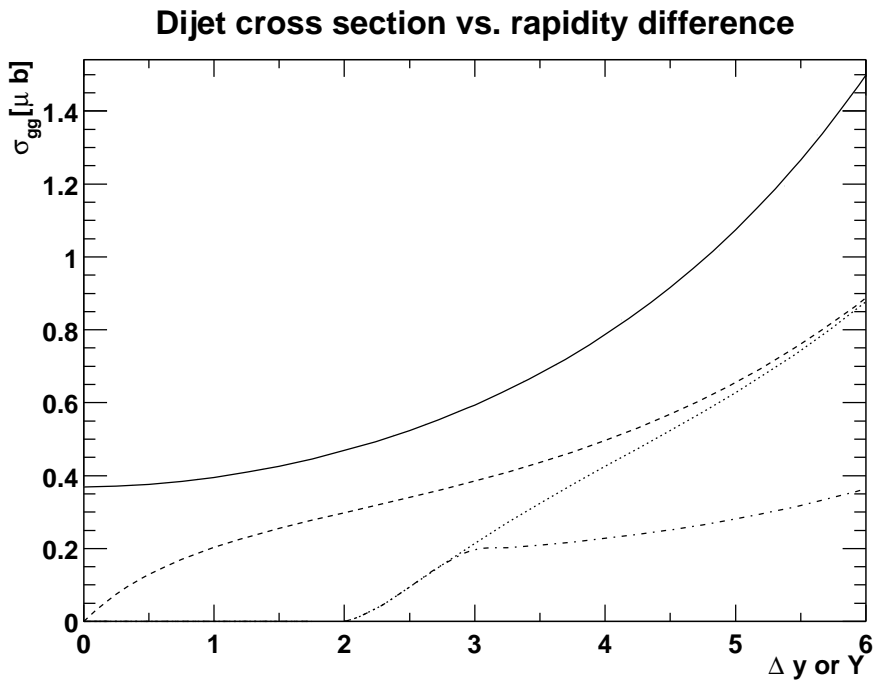,width=13cm} } \caption{
    \label{d0ggxsec} The dependence of the LL BFKL gluon-gluon cross section
    on $\Delta y$ in the standard Mueller-Navelet calculation
    (upper solid line) and on $Y$ for the D0 setup.
    Three curves are shown for the definition of $x$'s
    applied in the D0 analysis: dashed line for the requirement $\Delta y>0$,
    dotted line for $\Delta y>2$, dash-dotted for $Q^2_{\rm max}$ of
    Eq.~(\ref{qmax2}). }
\end{figure}

In dijet production at hadron colliders, at large enough rapidities, 
the rapidity interval is well approximated by the expression
$\dy\simeq \ln(\hs/|\hat t|)$, where $\hs=x_a x_b S$ and 
$|\hat t|\simeq \kta\ktb$, with $k_{a,b\perp}$ being the moduli of
the transverse momenta of the two jets,
$x_{a,b}$ the momentum fractions of the incoming partons,
and $\sqrt{S}$ the hadronic centre-of-mass energy. 
Once the transverse momenta are fixed, there are two ways of increasing
$\dy$: by increasing the $x$'s in a fixed energy 
collider; or viceversa, by fixing the $x$'s and letting $S$ grow, in a
ramping run collider experiment. The former set-up, the only feasible at
a collider run at fixed energy, has been proven to be unpractical, 
since in the dijet production rate $d\sigma/d\dy$ as a function of $\dy$
it is difficult to disentangle the BFKL-driven rise of 
the parton cross section from the steep fall-off of the parton 
densities~\cite{DelDuca:1994mn}. The latter set-up,
even though the first to be proposed~\cite{Mueller:1987ey}, has been 
analysed only lately~\cite{Abbott:2000ai}, because 
it required a collider running at different centre-of-mass energies.
Here we review first the original Mueller-Navelet 
proposal~\cite{Mueller:1987ey}, and
then analyse its implementation.\\
In the high-energy limit, $\hat s\gg |\hat t|$, 
any QCD scattering process is dominated by gluon exchange in the crossed 
channel. This constitutes the leading term of the BFKL resummation.
The corresponding QCD amplitude
factorizes into an effective amplitude formed by 
two scattering centres, the impact factors, connected by the gluon 
exchanged in the crossed channel. The BFKL equation then resums the
leading logarithmic
(LL) corrections, of ${\cal O}(\as^n\ln^n(\hat s/|\hat t|))$,
to the gluon exchange in the crossed channel. 
In dijet production at large rapidity intervals,
one can write the cross section in the following 
factorized form~\cite{Mueller:1987ey}
\beq
{d\sigma\over dx_ a^0 dx_b^0} = \int d\ktat d\ktbt
f_{\rm eff}(x_a^0,\muft)\, f_{\rm eff}(x_b^0,\muft)\,
{d\hat\sigma_{gg}\over d\ktat d\ktbt}\, ,\label{mrfac}
\eeq
where $\muf$ is the factorization scale, and
the effective parton distribution functions (p.d.f.)
are~\cite{Combridge:1984jn} $f_{\rm eff}(x) = G(x) + {4\over 9}\sum_f
\left[Q_f(x) + \bar Q_f(x)\right]$,
where the sum runs over the quark flavours, and we understand
a dependence of the p.d.f. also on the factorization scale.
$x_a^0,\, x_b^0$ are the 
parton momentum fractions in the high-energy limit,  
\beq
x_a^0 = \frac{\kta}{\sqrt{S}} e^{y_a}\qquad \qquad
x_b^0 = \frac{\ktb}{\sqrt{S}} e^{-y_b}\, ,
\label{nkin0}
\eeq
where $y_a$ ($y_b$) is the rapidity of the most forward (backward) jet.
In the high-energy limit, the gluon-gluon scattering cross section 
becomes~\cite{Mueller:1987ey}
\beq
{d\hat\sigma_{gg}\over d\ktat d\ktbt}\ =\
\biggl[{3\as\over \ktat}\biggr] \,
f(-\ktat,\ktbt,\dy) \,
\biggl[{3\as\over \ktbt}\biggr] \ .\label{cross}
\eeq
with $\dy = y_a-y_b \ge 0$. The quantities in square brackets 
are the impact factors for jet production. 
The function $f$ is the solution of the BFKL resummation.
Then one substitutes \eqn{cross} into
Eq.~(\ref{mrfac}) and integrates it over the transverse momenta $\kta$ and 
$\ktb$ above
a threshold $\Ecut$, at fixed coupling $\as$ and fixed $x_a^0,\, x_b^0$. 
The rapidity interval $\dy$ is determined from the
$x$'s (\ref{nkin0}),
\beq
\dy = \ln{x_a^0 x_b^0 S\over \kta\ktb} \label{rapint}
\eeq
and since it depends on $\kta\ktb$, it is not a constant 
within the integral. However, the dominant
contribution to \eqn{cross} comes from the largest value
of $\dy$, which is attained at the transverse momentum threshold, thus
in Ref.~\cite{Mueller:1987ey} $\dy$ is fixed at its maximum by reconstructing 
the $x$'s at the kinematic threshold for jet production and setting them in a 
one-to-one correspondence with the jet rapidities
\beq
x_a^\mn = \frac{\Ecut}{\sqrt{S}} e^{y_a}\qquad \qquad
x_b^\mn = \frac{\Ecut}{\sqrt{S}} e^{-y_b}\, .
\label{mnkin}
\eeq
Then the factorization formula (\ref{mrfac}) is determined at fixed 
$x_a^\mn, x_b^\mn$, and the integration over $\kta$ 
and $\ktb$ can be straightforwardly performed.
Asymptotically, the gluon-gluon cross section becomes
\beq
\hat\sigma_{gg}^{(\dy\gg 1)}(k_{a,b\perp}\!>\!\Ecut)\, = 
{9\pi\as^2\over 2 \Ecutt}
{e^{A \dy}\over \sqrt{\pi B \dy/4}}\, ,\label{asympsol}
\eeq
with $A = 12\ln{2}\as/\pi$ and $B = 42\zeta(3)\as/\pi$.
At very large rapidities the resummed gluon-gluon cross section
grows exponentially with $\dy$ (\fig{d0ggxsec}, solid line), 
in contrast to the LO (${\cal O}(\alpha_s^2)$) cross section, which is 
constant at large $\dy$.
In \eqn{mrfac}, the parton momentum fractions being basically fixed, 
one can interpret a rise of the jet cross section as a function
of $\dy$ as due to a rise in the parton cross section.
Then from the asymptotic formula~(\ref{asympsol}) the effective BFKL 
intercept \mbox{$\alpha_{\sss BFKL}\equiv A+1$} can be derived. 

The D0 Collaboration~\cite{Abbott:2000ai} has revisited 
the original Mueller-Navelet proposal, and has measured the ratio
\beq
R = {\sigma(\sqrt{S_{_A}}) \over \sigma(\sqrt{S_{_B}})}
\label{ratio}
\eeq
of dijet cross sections obtained at two different centre-of-mass
energies, $\sqrt{S_{_A}} = 1800$ GeV and $\sqrt{S_{_B}} = 630$ GeV.
The dijet events have been selected by tagging the most forward/backward 
jets in the event, and the cross section is measured as a function
of the momentum transfer, defined as $Q^2=\kta\ktb$, and of the
quantities
\beq
x_1 = \frac{2\kta}{\sqrt{S}} e^{\bar y} \cosh (\dy/2)\, ,\qquad
x_2 = \frac{2\ktb}{\sqrt{S}} e^{-\bar y} \cosh (\dy/2)\, ,
\label{eq:one}
\eeq
with $\bar y = (y_a+y_b)/2$. $x_1$ and $x_2$ 
are reconstructed from the tagged jets using~\eqn{eq:one}, irrespective 
of the number of additional jets in the final state. In 
LO kinematics, $x_1$ and $x_2$ are the momentum fractions
of the incoming partons. Higher-order corrections imply that 
these equalities no longer hold; however, \eqn{eq:one} is still a
reasonable approximation, except near the borders of  
phase space.  Thus when the ratio in Eq.~(\ref{ratio})
is computed at fixed $x_1$ and $x_2$, the contributions due to the p.d.f.'s
cancel to a large extent, allowing so the study of
BFKL effects without any contamination from long-distance phenomena.
Then, if \eqn{asympsol} holds, the ratio (\ref{ratio}) allows a determination
of the BFKL intercept.

In the analysis performed by D0~\cite{Abbott:2000ai}, jets have been
selected by requiring $k_{a,b\perp}>20$~GeV, $|y_{a,b}|<3$, and $\dy>2$,
and a cut on the momentum transfer, \mbox{$400<Q^2<1000$~GeV$^2$}, 
has been imposed. These cuts select dijet events at large rapidity intervals.

In the D0 setup there are, however, two main differences 
with respect to the standard Mueller-Navelet
analysis, in which it is assumed that the $x$'s are reconstructed through
\eqn{mnkin} and that the jet transverse momenta are unbounded from above:
\begin{itemize}
\item[$a)$] D0 collect data with an upper bound on $Q^2=\kta\ktb$, which is
of the same order of magnitude as the square of the lower cut on the jet 
transverse momenta, and thus cannot be ignored in the integration
over the transverse momenta;
\item[$b)$]D0 reconstruct the $x$'s through \eqn{eq:one}, which is well
approximated by \eqn{nkin0}, but not by \eqn{mnkin}.
\end{itemize}
Following Ref.~\cite{Andersen:2001kt},
we examine the consequences of these differences on the rise
of the cross section. First, we note that when we use \eqn{nkin0}, the jet 
rapidities are not fixed, rather in a given ($x_a^0,\,x_b^0$) bin all the 
transverse momenta and rapidities contribute which fulfil \eqn{nkin0}.
Thus the rapidity interval between the jets cannot be used as an independent,
fixed observable.
For convenience, we write the rapidity interval (\ref{rapint}) as 
\beq
\dy = Y + \ln{\Ecutt\over \kta\ktb}\, ,\label{uno}
\eeq
with $Y = \ln(x_a^0x_b^0 S/\Ecutt)$.
The requirement that the rapidity interval be positive, $\dy\ge 0$,
imposes an effective upper bound on $Q^2$: $Q^2_{\rm max} = \Ecutt e^Y$.
Integrating then the
gluon-gluon cross section~(\ref{cross}) over $\kta$ and $\ktb$ above 
$\Ecut$, at fixed $x_a^0,\, x_b^0$ and fixed coupling $\alpha_s$,
we obtain the dashed line of \fig{d0ggxsec}.
Note that as $Y\to 0$, the upper bound on $Q^2$ goes to the kinematic 
threshold, $Q^2_{\rm max} \to\Ecutt$, and accordingly the cross section 
vanishes. When we include the D0 
experimental cuts on the transverse momentum, $Q^2<1000$ GeV$^2$, and 
the rapidity interval, $\dy\ge 2$, the same analysis can be repeated 
except that the upper bound on $Q^2$ is given by
\beq
Q^2_{\rm max} = {\rm min}(1000\, {\rm GeV}^2,\,\Ecutt e^{(Y-2)})\, 
,\label{qmax2}
\eeq
where we have used the fact that $\Delta y>2$ imposes the second
effective upper bound on $Q^2$.
Integrating then the gluon-gluon cross section~(\ref{cross}) over 
$\kta$ and $\ktb$, we obtain the dot-dashed line of \fig{d0ggxsec}.
Thus the shape of the cross section as a function of $Y$ depends 
crucially on the upper bound on $Q^2$.
Asymptotically~\cite{Andersen:2001kt}, all of the curves of \fig{d0ggxsec}
have the same scaling with $Y$ as \eqn{asympsol} does. The fact that
this is not what we see in \fig{d0ggxsec}, signals that within
the D0 kinematic regime we are far from the asymptotic region.
Care must therefore be taken in interpreting any observed cross section
increase in the D0 data as due exclusively to the BFKL intercept.

In addition, the analysis of dijet production according to the D0 
acceptance cuts through the NLO partonic event generator of 
Ref.~\cite{Frixione:1997np} showed the presence of large logarithms of 
non-BFKL origin~\cite{Andersen:2001kt}, 
due to the choice of equal transverse momentum
cuts, $\kta,\ktb > \Ecut$. Such logarithms are bound to affect the BFKL
analysis, but can be easily avoided if different cuts on the transverse
momenta of the two jets are chosen. Accordingly, predictions are given
in Ref.~\cite{Andersen:2001kt} for the D0 ratio (\ref{ratio}), 
using LO and NLO QCD,
standard (``naive'') BFKL, and BFKL with energy-momentum conservation
and running coupling effects
through a partonic event generator~\cite{Schmidt:1997fg,Orr:1997im}.
Such predictions could be compared to the D0 data, if these are re-analysed
with different transverse momentum cuts on the tagging jets.

\section{Forward $W$ boson production with associated jets}

In the previous section, we have reviewed the feasibility of extracting
footprints of the BFKL resummation through the Mueller-Navelet analysis of 
forward dijet production; we mentioned, but dismissed summarily, 
dijet production
at a fixed energy collider because an eventual BFKL-driven rise of the
parton cross section within the rate $d\sigma/d\dy$, as a function of
$\dy$, would be hindered by the steep fall-off of the p.d.f.'s.
Infact, except at large $x$'s (i.e. $x \gsim 0.1$),
dijet production is dominated by gluon-gluon scattering, and since
the shape of the gluon p.d.f. is very sensitive to
$x$ variations (and monotonically decreasing with it), the dijet production
rate turns out to be sensitive to the approximation made (e.g. in
the BFKL analysis) in reconstructing the $x$'s from the jet kinematic
variables.

\begin{figure}[htb]
\begin{center}
\epsfig{file=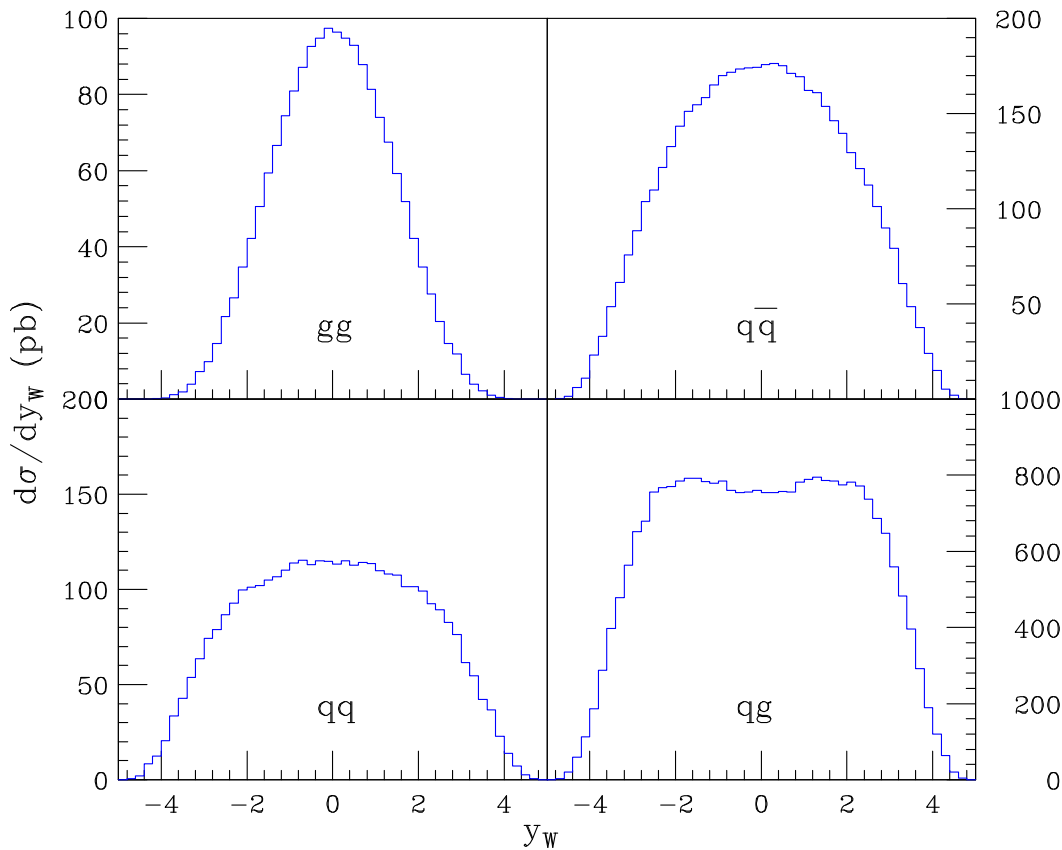,width=14cm}
\caption{Distributions in $\yw$ for the subprocesses of \eqn{subpro} at
the LHC centre-of-mass energy $\sqrt{s} =$ 14 TeV and with
$p_{j_\perp{\rm min}}=$ 30 GeV.}
\label{fig:w2jetyw}
\end{center}
\end{figure}

In this section, following the analysis of Ref. \cite{Andersen:2001ja}, 
we shall revive the quest for BFKL footprints
at a fixed energy hadron collider by considering forward $W$ boson production
in association with two jets. We believe to have reasons to
prefer forward $W$ boson production in association with two jets to
dijet production. Firstly, it might be
experimentally easier to pick up forward $W$ bosons that decay
leptonically than forward jets; once a forward lepton has triggered the event,
one observes the jets that are associated to it, with no limitations 
on their
transverse energy. Conversely, in a pure jet sample one usually triggers the
event on a jet of relatively high transverse energy, thus the triggering jet
cannot be too forward.
Secondly, $W$ production in association with jets lends
itself naturally to extensions to the high-energy limit, since it favours
configurations with a forward $W$ boson.
Presently we examine this remark by analysing in detail the
kinematics of $W + 2$ jet hadroproduction.
At LO the parton subprocesses are
\beqa
&(a)& g\, g\to W\, q\, \bar{q}\, ,\nonumber\\
&(b)& q\, \bar{q}\to W\, g\, g + W\, q\, {\bar q}\, ,\nonumber\\
&(c)& q\, q\to W\, q\, q\, ,\nonumber\\
&(d)& q\, g\to W\, q\, g\, .\label{subpro}
\eeqa
The momentum fractions of the incoming partons are given through
energy-momentum conservation by
\beqa
x_a &=& {k_{j_{1\perp}}\over\sqrt{s} } e^{y_{j_1}} +
{k_{j_{2\perp}}\over\sqrt{s} } e^{y_{j_2}} +
{m_\perp\over\sqrt{s} } e^{\yw}
\label{w2jkin}\\
x_b &=& {k_{j_{1\perp}}\over\sqrt{s} } e^{-y_{j_1}} +
{k_{j_{2\perp}}\over\sqrt{s} } e^{-y_{j_2}} +
{m_\perp\over\sqrt{s} } e^{-\yw} \nonumber
\eeqa
with $k_{j_{1,2\perp}}$ the jet transverse momenta and
$m_\perp = \sqrt{\mw^2 + |k_{j_{1\perp}}+k_{j_{2\perp}}|^2}$ the
$W$ transverse mass.

In \fig{fig:w2jetyw} we plot the rapidity distribution of the $W$ boson
for the four subprocesses considered above. 
$q\, g\to W\, q\, g$ is numerically dominant over the others.
$g\, g\to W\, q\, \bar{q}$ is perfectly symmetric,
thus the $W$ boson and the two jets are produced mostly in the central
rapidity region. However, in the other subprocesses
\begin{figure}[!hbt]
\begin{center}
\epsfig{file=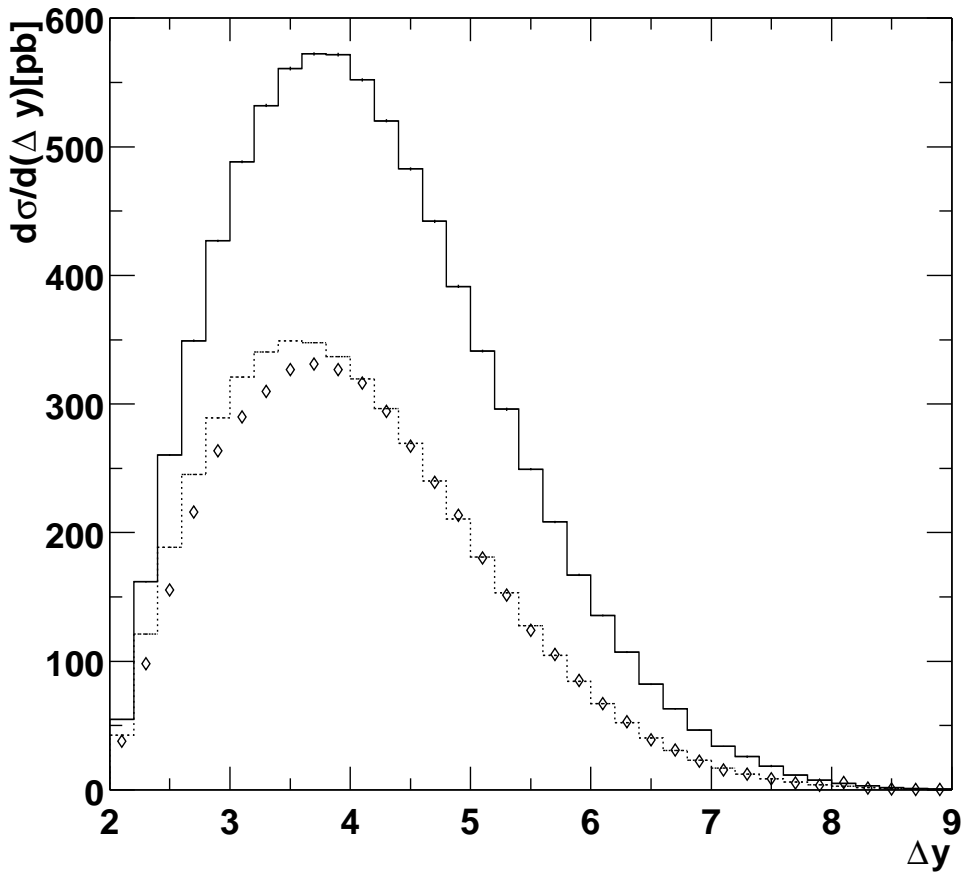,width=14cm}
\caption{The $W+2$-jet production rate as a function of the
rapidity interval between the jets $\Delta y = |y_{j_1}-y_{j_2}|$,
with acceptance cuts $y_{\sss W},\, y_{j_2} \ge 1$ and $y_{j_1} \le -1$,
or $y_{\sss W},\, y_{j_2} \le -1$ and $y_{j_1} \ge 1$.
The diamonds are the exact production rate;
the dashed-dotted curve is the production rate in the high-energy limit;
the solid curve includes the BFKL corrections and energy-momentum 
conservation through a BFKL partonic event generator.}
\label{fig:ywmy1}
\end{center}
\end{figure}
that is not the case: as
we move from $(a)$ to $(d)$ the $W$ boson tends to be produced
more and more forward in rapidity.
As in the $W^\pm$ rapidity asymmetry and in $W + 1$ jet production,
the physical mechanism is the
difference in the shape of the p.d.f.'s of the incoming partons.
In addition, 
one jet, say $j_2$, is always linked to the $W$ boson via a quark
propagator, as in $W+1$-jet production, so it tends to follow the $W$
in rapidity. The position
of the other jet is a dynamical feature peculiar of $W+2$-jet production:
thanks to the gluon exchanged in the crossed channel, 
in $(b)$, $(c)$ and $(d)$ that
jet can be easily separated in rapidity from the $W$ boson.

In \fig{fig:ywmy1} we consider $W+2$-jet production as a function of $\Delta
y$, and with acceptance cuts $y_{\sss W},\, y_{j_2} \ge 1$ and $y_{j_1} \le
-1$, or $y_{\sss W},\, y_{j_2} \le -1$ and $y_{j_1} \ge 1$, namely
we put a veto on tagging jets in the central rapidity region.
Note that \fig{fig:ywmy1}
is both qualitatively and quantitatively different from $d\sigma/ d\Delta y$
in dijet production~\cite{DelDuca:1994mn}: the peak is a striking
confirmation of the dominance of the configurations asymmetric in rapidity,
discussed above.  In fact the veto in the central rapidity
region strongly penalises the asymmetric configurations when $\Delta y$
approaches its minimum value; since the asymmetric configurations dominate
the $W+2$-jet production rate, the effect is a strong depletion of the
latter. In addition, the BFKL ladder, which includes
energy-momentum conservation, shows a substantial increase of
the cross section with respect to the LO analysis, as
opposed to the decrease of the dijet case~\cite{Orr:1997im}. Infact,
the implementation of energy-momentum conservation in the 
BFKL partonic event generator has a much lesser
impact than in the dijet case.  This is due to the fact that the valence 
quark distribution in $q\, g\to q\, g\, W$ is much less sensitive to $x$ 
variations than the gluon distribution in $g\, g\to g\, g$.
Secondly, the presence of at least three particles in the final state
makes the threshold configurations, and thus the logarithms of non-BFKL
origin, much less compelling than in the dijet case.

The analysis above lets us hope that $W + 2$ jet production
at the LHC will turn out to be a good test in favour of, or against, the
presence of the BFKL resummation in hadron collisions.

\end{document}